\documentclass[draftcls,12pt,onecolumn,oneside]{IEEEtran}
\usepackage{mathrsfs}
\usepackage{url,times,amsmath,color,amssymb,graphicx,epsfig,psfig,cite,geometry,psfrag,subfigure,dsfont,booktabs}
\usepackage{algorithm}
\usepackage{algorithmic}
\usepackage{bm}
\begin{document}
\title{Resource Allocation in NOMA based Fog Radio Access Networks}
\author{Haijun Zhang,~\IEEEmembership{Senior Member,~IEEE}, Yu Qiu,  \\
  Keping Long,~\IEEEmembership{Senior Member,~IEEE},  George K. Karagiannidis,~\IEEEmembership{Fellow,~IEEE}, Xianbin Wang,~\IEEEmembership{Fellow,~IEEE}, and Arumugam Nallanathan,~\IEEEmembership{Fellow,~IEEE}
\thanks{Haijun Zhang and  Keping Long are with the Engineering and Technology Research Center for Convergence Networks, University of Science and Technology Beijing, Beijing, 100083, China (e-mail: haijunzhang@ieee.org, longkeping@ustb.edu.cn).

Yu Qiu is with College of Information Science and Technology, Beijing University of Chemical Technology, Beijing, 100029, China (e-mail: eeqiuyu@gmail.com).

George K. Karagiannidis is with Aristotle University of Thessaloniki, Thessaloniki , Greece (e-mail: geokarag@auth.gr).

Xianbin Wang is with University of Western Ontario, London, Canada (e-mail: xianbin.wang@uwo.ca).

Arumugam Nallanathan is with Queen Mary University of London, United Kingdom (e-mail: nallanathan@ieee.org).

}}
\maketitle
\begin{abstract}
In the wake of growth in intelligent mobile devices and wide usage of bandwidth-hungry applications of mobile Internet, the demand of wireless data traffic and ubiquitous mobile broadband is rapidly increasing. On account of these developments, the research on fifth generation (5G) networks presents an accelerative tendency on a global scale. Edge computing draw lots of attention for reducing the time delay and improving the Quality of Service for the networks. While, fog radio access networks (F-RANs) is an emergent architecture, which takes full use of edge computing and distributed storing capabilities in edge devices. In this article, we propose an architecture of non-orthogonal multiple access (NOMA) based F-RANs, which has a strong capability of edge computing and can meet the heterogeneous requirements in 5G systems. NOMA with successive interference cancellation (SIC) is regarded as a critical multi-user access technology. In NOMA, more than one user can access the same time, code domain, and frequency resources. With assigning different power levels to multi-user and implementing SIC, multiple users detection can be achieved. In this article, we provide a description of the NOMA based F-RANs architecture, and discuss the resource allocation in that. We will focus on the power and subchannel allocation in consideration of using NOMA and the edge caching. Simulation results show that the proposed NOMA baesd F-RANs architecture and the resource management mechanisms can achieve the high net utility for the RANs.
\end{abstract}

\section{Introduction}

With the increasing popularity of new applications and the billions of mobile users, the past few decades have witnessed a enormous growth of mobile and wireless networks. More intelligent phones, connected vehicles, and other smart Internet of Things devices require seamless and stable network connectivity, which results the huge amount of traffic data. Meanwhile, this development presents the growing demand for broadband spectrum, high scalability, ultra-low latency, less consumption. In order to satisfy these requirements, many global research centres have proposed some novel technologies. During the past few years, cloud radio access networks (C-RANs) has drawn much attention and been regarded as a promise approach to handle the tremendous amount of devices \cite{SC2015}. In C-RANs, a crowd of Remote Radio Heads (RRHs) are distributed within a particular geographical region, which are connected to a concentrated BBU pool via the high bandwidth fronthaul links. Coordination among BBU pools improve resource utilization and reduce power consumption. While, the fronthaul is commonly capacity and latency constrained in actually, leading to dramatic reductions in spectral efficiency and energy efficiency performance of networks.


To compensate the drawbacks of C-RANs, a new paradigm called Fog Computing has been introduced \cite{FB2012}. Cisco has proposed the concept of F-RANs, which integrates the fog computing with radio access network. Unlike the apparent centrality of C-RANs, F-RANs extend a high proportion of functions to the edge of the network, such as computing, storage, control, management, application services and so on \cite{HY2017}. Specifically, a great deal of signal processing and computing is performed in a distributed manner, and some data can be stored in the edge devices. Moreover, the real time collaboration radio signal processing and flexible cooperative radio resource management can be executed in the edge of network. In F-RANs, the fog-computing access point (F-APs) and fog user equipments (F-UEs) are equipped with caching. Besides, the F-APs can execute radio signal processing locally using their adequate computing capabilities and can manage their caching memories flexibly. These distinct features of F-RANs make a contribution to provide the superior user experience and improve network performance, including the high mobility and ultra low latency.

Meanwhile, NOMA has been validated as an promising multiple access mechanisms for future RANs to meet the heterogeneous demands for low latency, massive connectivity, and high throughput \cite{PX2015}. By utilizing the power domain rather than the conventional time and frequency domains, the NOMA can significantly improve the network throughput. The essential idea of NOMA is that multiple users can share the same frequency resources and use different power levels \cite{FF2017}. Meanwhile, there are inevitable expense of inter-user interference. SIC is applied as a effective multiple users detection technology. Using NOMA with SIC, users with better channel conditions first successively subtracts the messages of users with worse channel conditions and remove the interference before decoding its own messages \cite{FF2016}. Therefore, NOMA with SIC can be employed in F-RANs to ensure multiple users with simultaneous transmissions in complex wireless environments. A many-to-many subchannel matching algorithm was proposed and approached the performance of the upper bound and greatly outperformed the OFDMA networks. Tree search scheme was introduced as a low computational complexity power assignment method for NOMA with SIC receiver in \cite{ZZ2014}. On the standpoint of energy aspects, the cooperative NOMA transmission protocol was proposed in \cite{YZ2016}, in which near NOMA users were regarded as energy harvesting user relays for forwarding messages to far NOMA users. To the best of the authors¡¯ knowledge, it is the first attempt to investigate the application of NOMA in a F-RANs environment and resource allocation with interference management in the literature.

In this article, we investigate a solution of resource allocation to improve the networks performance of NOMA based F-RANs. Firstly, we propose the network architecture of NOMA baesd F-RANs. Then, we study the resource management, which includes the power and subchannel allocation. The power allocation problem is modeled as a non-cooperative game. For subchannel allocation, we study the many-to-many two-sided matching algorithm. Moreover, we will show that the proposed resource management mechanisms enhance the net utility of the NOMA baesd F-RANs.

\section{System Architecture}

In Fig. 1, we present the architecture of NOMA based F-RANs. In this proposed networks, the functions of the control plane are completed in the marco remote radio heads (MRRHs), rather than the BBU pool in C-RANs. The BBU pool will mainly provide centralized storage and communications in NOMA based F-RANs. And the MRRHs are interfaced to BBU pool through backhaul links. All F-APs are connected to BBU pool by the fronthaul links. The F-APs are the evolution from traditional RRHs, which play the important role in NOMA based F-RANs. Unlike the centralized data storage in C-RANs, a substantial part of data are distributed in F-APs and some F-UEs as edge caching. Not only a certain amount of caching are assigned to F-APs, but also radio signal processing and radio resource management can be performed locally in F-APs. It is quite possible that F-UEs can directly connect the F-APs to get the desired content, instead of establishing complex transmission link with core network. Similarly, F-UEs can download the data from the neighboring F-UEs through the device-to-device (D2D) technology. And when the distance of two possible paired F-UEs is so far that they can not directly connect with each others, the third F-UE will play the role as relay to forwards the information.

\begin{figure}[t]
        \centering
        \includegraphics*[width=13cm]{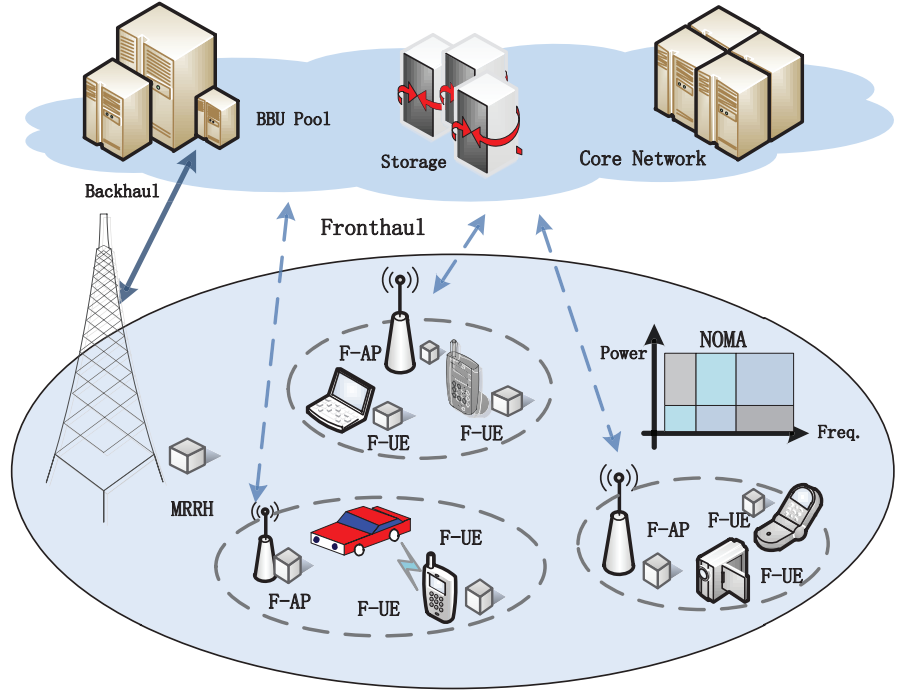}
        \caption{A NOMA based F-RANs architecture.}
        \label{fig:1}
\end{figure}

\subsection{Edge Caching Utilization}

As above, there are substantial data as the cache in the edge of network, such as F-APs and F-UEs. Noted that what is stored in the F-APs and F-UEs is not arbitrarily assigned. How to choose appropriate content as the edge caching has a profound effect on network performance, especially for time delay and energy consumption. With the upward popular tendency of location-based mobile applications, the enormous amounts of information may emerge anytime and anywhere. If all these surging data stream are transmitted to the centralized BBU pool, it will pushes the fronthaul links to their capacity limits. And it is found that there are quite great relevance of these data flow among F-UEs in close physical proximity. At the same time, some social applications makes the close-range F-UEs exchanging data traffic frequently than other distant F-UEs. Besides, F-UEs from the same social network or enjoying the same social activity can require the same data over the downlink. The state of network applications inspires us to effectively set cache placement and use the finite capacity of edge caching. Actually, F-APs and F-UEs store the most popular or relevant contents as edge caching until there is no space to stored, instead of the contents randomly assigned with equal probabilities. In these cases, the requested services can be completed locally by the edge caching of the popular contents. Consequently, F-UEs have no need to interface with the BBU pool every time when they request data. Similarly, if a crowd of F-UEs have the highly relevant data to be recorded in the cloud, there is no need for all F-UEs to upload traffic individually. Then, the contents of their common interests will be storaged as edge caching in only one single F-UE. Benefitting from edge caching, the amount of system data supported by both the fronthaul and backhaul links can be significantly decreased. These caching utilization in NOMA based F-RANs can reduce the long transmission time latency and heavy burden on the fronthaul and the BBU pool, which are the two hardest issues in both C-RANs.

\begin{figure}[t]
        \centering
        \includegraphics*[width=13cm]{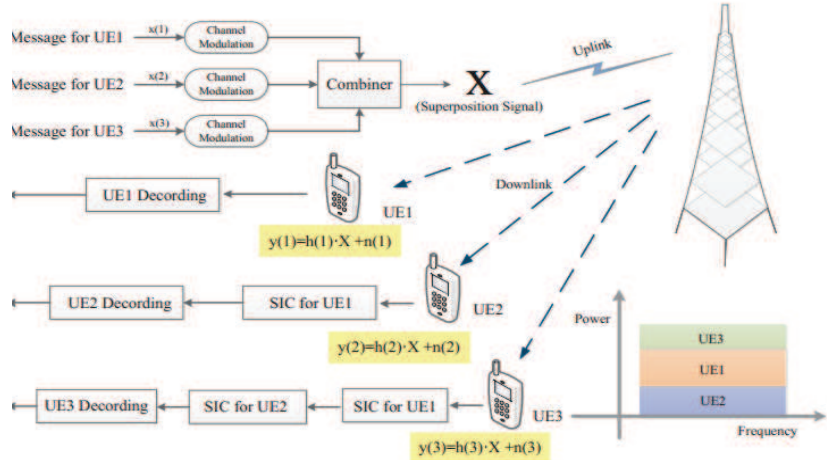}
        \caption{Basic NOMA scheme for three users. In downlink, SIC technology is applied to the user receivers. Three users are allocated on the same subchannel with the channel response normalized by noise (CRNN) order: ${{CNRR}_{3}}$ $\ge$ ${{CNRR}_{2}}$ $\ge$ ${{CNRR}_{1}}$.}
        \label{fig:1}
\end{figure}

\subsection{Noma Protocol}

Figure 2 shows the typical NOMA scheme for the scenario of three users. There are three users which are allocated on the same subchannel: ${x_1}(t)$, ${x_2}(t)$, and ${x_3}(t)$. Moreover, ${x_1}(t) > {x_2}(t) > {x_3}(t)$. The superposition signal at SIC receiver can be denoted as the $X$, including all users signals about channel interference and the additive white gaussian noise. Before the first stage of user detection, the SIC receiver will put these users in the increasing order of the channel response normalized by noise (CRNN). Consider that three users share the same subchannel with CRNNs order: $CRN{N_1} \le CRN{N_2} \le CRN{N_3}$. Depending on this order, user 1 with the smallest CRNN can firstly decode form the superposition signal. Then, the interference form the user 1 in the poorest channel condition can be correctly removed by the user 2 in better channel condition. Similarly, the user 3 can successively subtracts the interference form the user 1 and user 2 and then obtains its own message. Based on the increasing order of CRNN, user 1, 2 and 3 can successfully decode and cancel the interference symbols in turn.

Furthermore, the performance of networks can be further improved by using the NOMA technology in the D2D communication scenario. If a group of F-UEs in D2D mode transmit various contents, multiple bandwidth channels would be occupied in general. However, NOMA makes it possible that these receivers are served simultaneously in only one channel. For example, three F-UEs are considered in a D2D group, and what they required are video, audio and text messages respectively. Suppose the video and audio F-UEs are with good channel conditions, they can receive the accurate signals and achieve high data rates by SIC technology. Specifically, the impact of the partners in a D2D group can be completely eliminated after two or three times. As to the F-UE requiring text, the service can be successfully provided just with small data rates. Hence the strong co-channel interference can be safely ignored. The integrating with NOMA significantly enhances the performance of systems, especially in both SE and massive connectivity.

\section{Resource Allocation in NOMA Based F-RANs}

With the growing popularity of big data and Internet of Things in fog computing scenario, the provided services and resources are becoming more complicated than ever before. What we should consider is not only the unique resource heterogeneity, but also the resource limitations, locality restrictions \cite{XW2014}, and dynamic nature of resource demand. Accordingly, how to efficiently allocate the heterogeneous resources to achieve the optimal network utility, edge caching \cite{XX2015} should be taken account of.


We focus on resource allocation in terms of maximizing the sum data rate of the system. In our consideration, all the F-APs have the full knowledge of channel state information, and share spectrum resources with the central MRRH. Noticed that the spectrum sharing framework will bring about both co-tier and cross-tier interference. With adopting the NOMA protocol \cite{YY2013, AA2014}, F-UE multiplexing in the power domain is exploited, which makes different F-UEs served with different power values. Meanwhile, owing to one subchannel can be occupied by a subset of F-UEs, the signal of a F-UE will cause interference to the other F-UEs allocated to the same subchannel. Therefore, the received signal at the F-UE includes the desired signal, the additive white gaussian noise (AWGN), co-tier interference (from the F-UEs in other F-APs reusing the same subchannel), cross-tier interference (from the MRRH) and the interference from the NOMA F-UEs served by the same F-AP.


In this article, power allocation and subchannel allocation are coupled with each other in terms of maximizing the sum data rate. Firstly, F-APs allocate the transmission power to the F-UEs over each subchannel, depending on the non-cooperation game scheme. Then, the subchannel allocation can be modeled as many-to-many two-sided matching problem, and solved by the matching theory, as will be discussed in detail in the next subsection.

\subsection{Power Allocation in NOMA based F-RANs}

In the existing literature, game theory has been extensively used for interference mitigation in networks \cite{HX2012}. We define the maximum achievable data rate of the F-UE as the optimization function of each F-UE in system. The optimization of resource management can be formulated as a non-cooperative resource allocation game. Accordingly, all F-UEs in each F-AP only depend on their interest to achieve their own greatest benefit, acting as the selfish players in competition to satisfy themselves. While, there are huge power consumption when every F-UE adopts their maximal transmit power. Thus, the optimization function of overall F-APs can be modeled as the maximization of object function under the constraint of the maximal transmit power of each F-UE.

\subsubsection{Pricing Function of Interference }

In consideration of the fundamental role of MRRH in providing ubiquitous connection to achieve seamless coverage, the MRRH should be strictly protected from the impact by F-APs deployment. An interference threshold will be proposed to represent the maximum tolerable interference level for MRRH causing by massed F-UEs. At the same time, this density of F-UEs would interfere with MRRH, naturally leading to a significant decrease in the total network throughput and affect the capability of MRRH. It is essential to propose the pricing function to constrain the interference to MRRH. This pricing function can be applied to manage the interference creating by F-UEs to MRRH, which is proportional to the transmit power of each F-UE.

\subsubsection{Reward Function of Edge Caching}

Similarly, the reward function corresponding to the a certain edge caching strategy in NOMA based F-RANs should be introduced to represent the benefit from the storage resources in the edge of networks. When the contents storaged in F-APs or F-UEs is requested, the use of edge caching can be alleviation of bandwidth and reduction of delay. In this article, we choose the alleviation of backhaul bandwidth as the reward of edge caching. By means of the coefficient, which is about the utilization of edge caching, we can compute the a certain degree of compensation from the finite edge caching in NOMA based F-RANs.

\subsubsection{Utility Function}

According to these assumption above, we propose an interference-aware resource power allocation scheme for the downlink of co-channel deployed F-APs. The optimization function under the proposed resource management scheme can be achieved by maximizing the net utility function of each F-UE, which is defined as the maximum achievable data rate of the F-UE minus the pricing function of the interference and adds the reward function of the edge caching offered from networks.

As the fixed points, Nash Equilibrium (NE) has been proven unique and existing on each individual subchannel in non-cooperative resource allocation game. The power allocated for each F-UE is restricted between zero and the maximum power. We know that there are serious cross-tier interference and co-tier interference, when F-UE want to optimize its utility by equipped with the maximum transmission power, making the strategy far from Pareto-optimality. Starting with the smallest available power value, we can update the transmission power of each F-UE successively through the designed iterative algorithm to find the optimal value.

\subsection{Subchannel Matching in NOMA based F-RANs}

To deal with the allocation issue between the F-UEs and the subchannels, we propose a approach depending on matching theory. First, we assume a set of $M$ F-UEs form $K$ F-APs, and the set of $N$ subchannels. The players in these two disjoint sets are rational and selfish, who intend to obtain the peak value of their own benefit. When ${{SC}_{n}}$  is assigned to ${{F-UE}_{k,m}}$, it means that the ${{SC}_{n}}$ and ${{F-UE}_{k,m}}$ are matched with each other.

For a given subchannel, F-UE $i$ desires to decode and remove the interference from F-UE $j$ from the superimposed signals by the SIC technique. It is noticed that the application of SIC technology may lead to a certain complexity. The complexity at the receiver will increase along with the growing number of F-UEs reusing each subchannel. Hence, we should make a reasonable constraint to limit the maximum number of F-UEs occupying the same subchannel at the same time. Thus, the decoding complexity at the receiver can be tolerated. And there are significant improvement in the hardware complexity and processing delay.

About each player, the player of the other set has different preference. Noted that the preferences of each player do not rely on the other players' behaviors. Then, to better demonstrate the dynamic matching process of all players how to compete with each other and make decisions, we propose the preference lists of the F-UEs ${{Pre}({F-UE})}$ and subchannels ${{Pre}({SC})}$ respectively. The solution of the matching game is an allocation strategy between F-UEs and subchannels, which is based on the preference lists to satisfies each players preference.

If ${{F-UE}_{k,m}}$ has preference for ${{SC}_{i}}$ rather than ${{SC}_{j}}$, it implies that ${{F-UE}_{k,m}}$ can achieve higher channel gain when assigned to ${{SC}_{i}}$ than to ${{SC}_{j}}$. Similarly, we define $p$ and ${{p}^{'}}$ as the pair of F-UEs. When ${{SC}_{n}}$ prefers the F-UEs in $q$ to F-UEs in ${{p}^{'}}$, the F-UEs in set $p$ can provide higher profit than F-UEs in set ${{p}^{'}}$ on ${{SC}_{n}}$. And the preference of each subchannel over different subset of the set of F-UEs can be proposed. On the basis of that lists, it is clearly which one set of F-UEs works with better performance than the other one. Some matching schemes of different characteristics of preferences have been studied for various scenarios in \cite{AR1992} \cite{SR2014}. In this article, we formulate the subchannel assignment problem as a many-to-many matching problem, depending on the the preference lists.

As to many-to-many two-sided matching, there are some key definition. a) As to each user, it can match with a subset of subchannel. b) As to each subchannel, a subset of F-UEs can be matched. c) The number of F-UEs can be assigned on each subchannel is limited to $q$. d) Matching $\mu$ is a mapping which assigns at least two different F-UEs (but less than or equal to $q$) to each subchannel and mutiple subchannel to each F-UEs.

It is noted that the formulated matching model is more complicated than the traditional matching models. As above, in our model any F-UEs of each set can be matched with any subset of the subchannel sets. Although the finite F-UEs can be multiplexed on the same subchannel, the quantity of potential matching pairs may be very large when the F-UEs set involves many members. Besides, we should considerate the correlative dependence of F-UEs combinations which are allocated in the same subchannel. Each subchannel should choose a ideal pair of F-UEs to match with, resulting the matching process more complicated even if we do not take the power allocation into account. Thus, a suboptimal matching scheme is proposed for subchannel assignment as follows for reducing the complexity.

In the matching procedure, at each round, each F-UE will send the matching request to its most preferred subchannel. Basing the list of subchannels ordered by decreasing channel gains, the ${{F-UE}_{k,m}}$ will find the first non-zero entry. Then the ${{F-UE}_{k,m}}$ will send matching request to the corresponding subchannel. When the number of assigend F-UEs on the subchannel is less than $q$, the subchannel can accept the matching request directly. While, if there are $q$ F-UEs allocated to the subchannel, we should compare the value of net utility provided by the different subset of F-UEs. This subchannel will accept the subset of F-UEs which satisfy maximum net utility, or else the request will the rejected. The whole matching process is repeated until no available F-UE is left to be allocated. Then, the assigned F-UE and the corresponding subchannels in the preference list are set to zero.

\subsection{Simulation results}

\begin{figure}[t]
        \centering
        \includegraphics*[width=13cm]{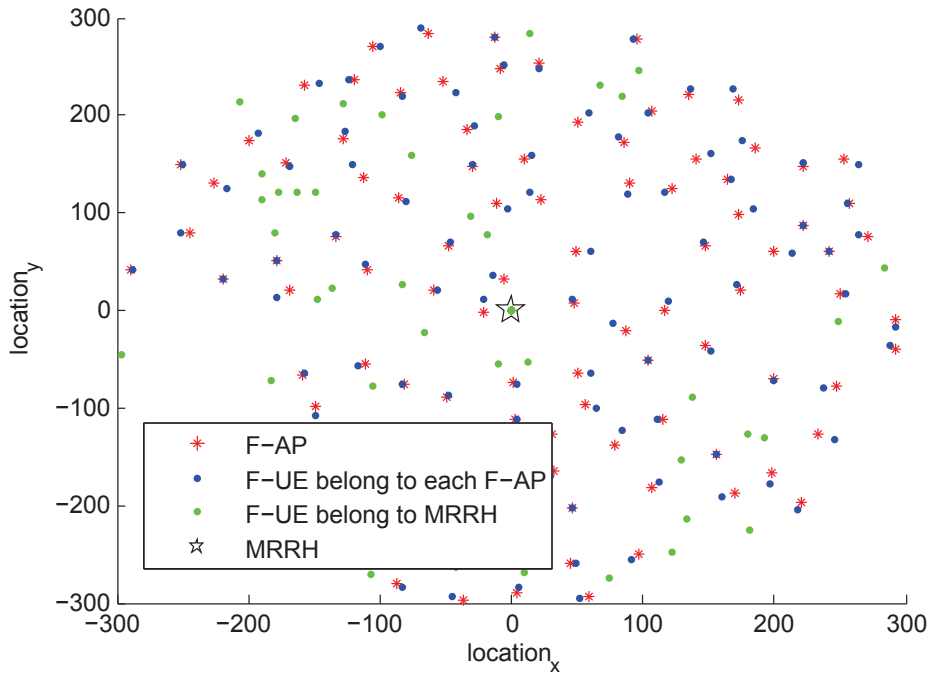}
        \caption{The system design of distance setting.}
        \label{fig:3}
\end{figure}

This section evaluates the performance of the proposed resource allocation algorithms for NOMA based F-RANs. In the simulations, the F-APs and F-UEs are randomly distributed in the central MRRH coverage area as illustrated in Fig. 3. We set the radius $R$ of MRRH and each F-AP to be 500 meters and 10 meters respectively. The minimum distance from MRRH to F-AP and F-UEs belong to MRRH is 300 m and 50 m respectively, and the minimum distance between F-APs is 40 m. The bandwidth is limited to 5 MHz. We set the peak power of each F-AP as 41 dBm, noise power spectral density as -174 dBm/Hz. While, in NOMA scheme, we restrict that only $q$ F-UEs are allocated to each subchannel for the sake of reducing demodulating complexity in the SIC receiver side.

 \begin{figure}[htbp]
        \centering
        \includegraphics*[width=13cm]{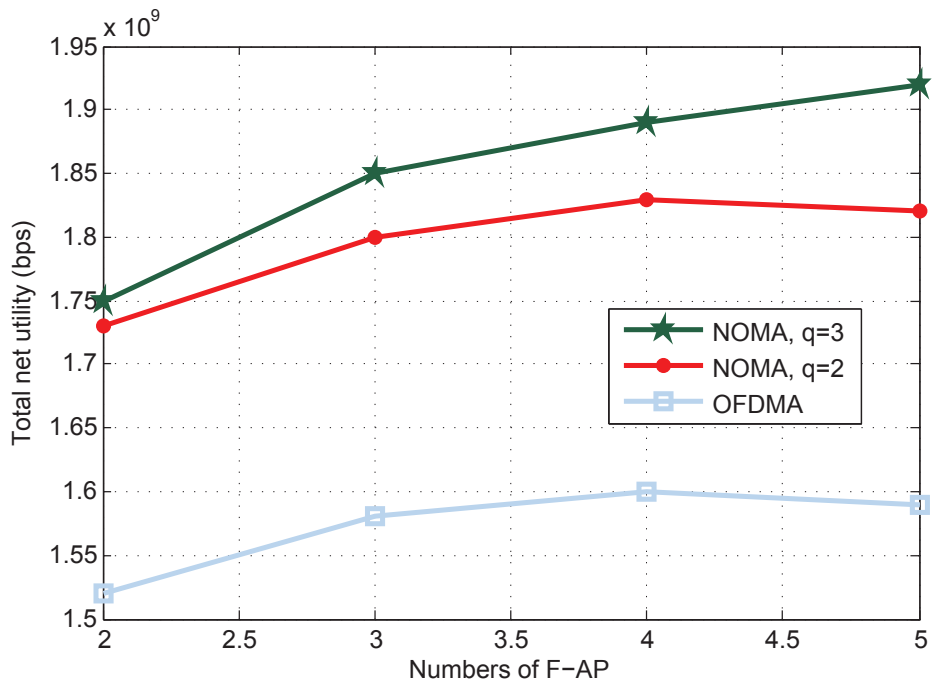}
        \caption{Total net utility of NOMA based FRAN versus different number of F-APs.}
        \label{fig:4}
\end{figure}

\begin{figure}[htbp]
        \centering
        \includegraphics*[width=12cm]{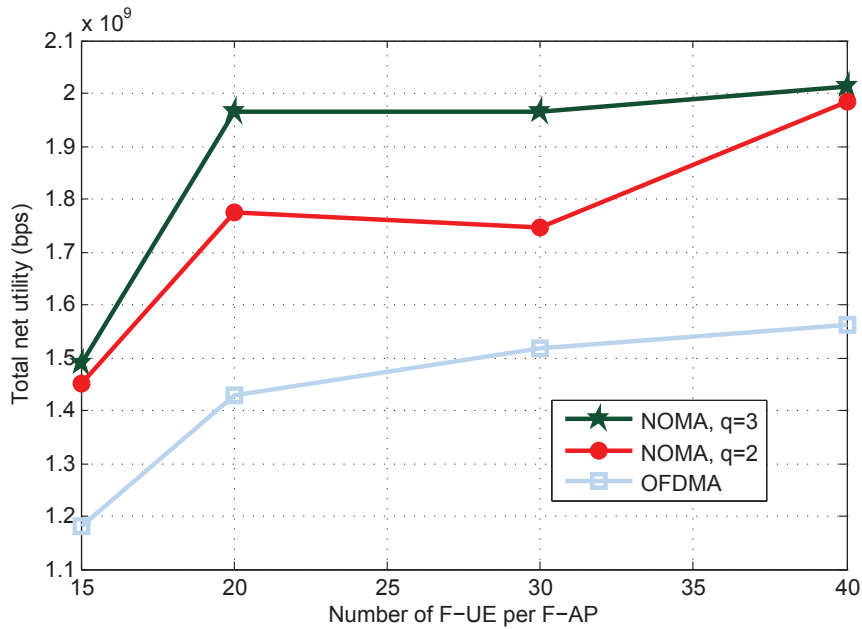}
        \caption{Total net utility of NOMA based FRAN versus different number of F-UE per F-AP.}
        \label{fig:5}
\end{figure}

In Fig. 4, the performance of NOMA based F-RANs is good with the increasing number of F-APs, and much higher net utility of system is achieved than that in OFDMA scheme. In other words, the scheme we proposed can guarantee good services, even in dense deployed networks. But there is little reduction when the number of F-APs from 4 to 5. It motivates us to find the optimal number of F-APs in the NOMA based F-RANs to achieve the best net utility. In Fig. 5, we compare the performance of the proposed scheme for NOMA based F-RANs with the trantional schemes of OFDMA, where $q$ is set as 2 and 3. As the number of the F-UE per F-AP grows, the net utility of NOMA is much higher than that of OFDMA owing to the multiuser diversity gain. For example, when the number of F-UE per F-AP is 30 and $q$ = 2, the net utility gain of the proposed resource allocation scheme is 29\% more than that of a conventional OFDMA scheme. This is because each subchannel can only be allocated to one F-UE in the OFDMA systems. In other words, the F-APs do not make the best of the finite frequency resources. And we will find that, when $q$ = 3, the proposed scheme performs better than this case in which $q$ = 2. This can be attributed to the fact that F-AP has more possible selections over the set of F-UEs to be allocated to each subchannel.

\section{Future Works}

However, there is still some room for further investigating in NOMA based F-RANs. Edge cache placement strategy should be seriously designed to provide the more flexible transmission opportunities to F-UEs. And the edge caching reward can be widely exploited in the content delivery process. While the edge caching space at each F-AP and F-UE are restricted and practically very small. At the same time, there are many crucial factors which should be taken into account and jointly optimized. For example, the cost of caching hardware makes a request to the hardware device about the storage capacity, which is the also increase the economic cost. Besides, the cache hit ratio and the number of data requests are play important roles in optimal radio resource usage.

Then, the trustworthiness and security is another key aspect we should investigated in futher. F-RANs can support the more flexible network operations in distributed manners. In consideration of the proximity to F-UEs and locality on the edge of networks, the massive nodes in F-RANs can often operate as the first node of access control. Thus, the networks edge nodes must be equipped with the right of monitoring and supervising the privacy-sensitive data. It is essential that we should take advantage of F-RANs to enhance network security instead of be restricted by its distributed characteristic. Meanwhile, multiple-input multiple-output, millimeter-wave communications, and Network Function Virtualization and other key technologies can be combined with NOMA based F-RANs to achieve better performance. In future we plan to work on solving some of these challenges.

\section{Conclusion}

In this article, the F-RANs architecture for 5G networks is proposed enhanced with NOMA. The NOMA based F-RANs architecture takes full advantage with the edge of networks. The power and subchannel allocation problems in NOMA based F-RANs were studied, with the aim of maximizing the net utility while considering the co-channel interference. For solving the power allocation problem, the resource optimization mechanisms is proposed under the non-cooperation framework. We present a suboptimal algorithm basing on formulating the subchannel allocation problem as a many-to-many two-side matching game. The numerical results have demonstrated that the NOMA based F-RANs architecture have more potential benefits in terms of net utility compared to conventional OFDMA networks.

\section*{Acknowledgment}
This work is supported by the National Natural Science Foundation of China (61471025, 61771044), the Young Elite Scientist Sponsorship Program by CAST (2016QNRC001), the Research Foundation of Ministry of Education of China \& China Mobile (MCM20170108), Beijing Natural Science Foundation (L172025), and the Fundamental Research Funds for the Central Universities(FRF-GF-17-A6, RC1631).

\begin{IEEEbiography}{Haijun Zhang} (M'13, SM'17) is currently a Full Professor in University of Science and Technology Beijing, China. He was a Postdoctoral Research Fellow in Department of Electrical and Computer Engineering, the University of British Columbia (UBC), Vancouver Campus, Canada. He serves as Editor of IEEE Transactions on Communications, IEEE 5G Tech Focus, and serves/served as a Leading Guest Editor for IEEE Communications Magazine, and IEEE Transactions on Emerging Topics in Computing. He serves/served as General Co-Chair of GameNets'16, Symposium Chair of Globecom'19, TPC Co-Chair of INFOCOM'18 Workshop IECCO, General Co-Chair of ICC'18/ICC'17/Globecom'17 Workshop on UDN, and General Co-Chair of Globecom'17 Workshop on LTE-U. He received the IEEE ComSoc Young Author Best Paper Award in 2017.
\end{IEEEbiography}

\begin{IEEEbiography}{Yu Qiu} received the BS degree in electronic information engineering from Beijing University of Chemical Technology, Beijing, China, in 2015. She is currently pursuing the M.S. degree at the Laboratory of Wireless Communications and Networks from College of Information Science \& Technology, Beijing University of Chemical Technology, Beijing, China. Her research interests include 5G networks, Network Functions Virtualization, mobility management and resource allocation in wireless communications.
\end{IEEEbiography}

\begin{IEEEbiography}{Keping Long}(SM'06) received his M.S. and Ph.D. degrees at UESTC in 1995 and 1998, respectively. He worked as an associate professor at BUPT. From July 2001 to November 2002, he was a research fellow in the ARC Special Research Centre for Ultra Broadband Information Networks (CUBIN) at the University of Melbourne, Australia. He is now a professor and dean at School of Computer and Communication Engineering (CCE), USTB. He is a member of the Editorial Committee of Sciences in China Series Fand China Communications. He is also a TPC and ISC member for COIN, IEEE IWCN, ICON, and APOC, and Organizing Co-Chair of of IWCMC'06, TPC Chair of COIN'05/'08, and TPC Co-Chair of COIN'08/'10, He was award-ed the National Science Fund Award for Distinguished Young Scholars of China in 2007 and selected as the Chang Jiang Scholars Program Professor of China in 2008. He has published over 200 papers, 20 keynotes, and invited talks.
\end{IEEEbiography}

\begin{IEEEbiography}{George K. Karagiannidis}(M'96-SM'03-F'14) was born in Pithagorion, Samos Island, Greece. He received the University Diploma (5 years) and PhD degree, both in electrical and computer engineering from the University of Patras, in 1987 and 1999, respectively. From 2000 to 2004, he was a Senior Researcher at the Institute for Space Applications and Remote Sensing, National Observatory of Athens, Greece. In June 2004, he joined the faculty of Aristotle University of Thessaloniki, Greece where he is currently Professor in the Electrical Computer Engineering Dept. and Director of Digital Telecommunications Systems and Networks Laboratory.  He is also Honorary Professor at South West Jiaotong University, Chengdu, China.

His research interests are in the broad area of Digital Communications Systems and Signal processing, with emphasis on Wireless Communications, Optical Wireless Communications, Wireless Power Transfer and Applications, Molecular and Nanoscale Communications, Stochastic Processes in Biology and Wireless Security.

He is the author or co-author of more than 450 technical papers published in scientific journals and presented at international conferences. He is also author of the Greek edition of a book on ¡°Telecommunications Systems¡± and co-author of the book ¡°Advanced Optical Wireless Communications Systems¡±, Cambridge Publications, 2012.

Dr. Karagiannidis has been involved as General Chair, Technical Program Chair and member of Technical Program Committees in several IEEE and non-IEEE conferences. In the past, he was Editor in IEEE Transactions on Communications, Senior Editor of IEEE Communications Letters, Editor of the EURASIP Journal of Wireless Communications Networks and several times Guest Editor in IEEE Selected Areas in Communications. From 2012 to 2015 he was the Editor-in Chief of IEEE Communications Letters. Dr. Karagiannidis is IEEE Fellow and one of the highly-cited authors across all areas of Electrical Engineering, recognized as 2015, 2016 and 2017 Web-of-Science Highly-Cited Researcher.
\end{IEEEbiography}

\begin{IEEEbiography}{Dr. Xianbin Wang}(S'98-M'99-SM'06-F'17) is a Professor and Canada Research Chair at Western University, Canada. He received his Ph.D. degree in electrical and computer engineering from National University of Singapore in 2001.

Prior to joining Western, he was with Communications Research Centre Canada (CRC) as a Research Scientist/Senior Research Scientist between July 2002 and Dec. 2007. From Jan. 2001 to July 2002, he was a system designer at STMicroelectronics, where he was responsible for the system design of DSL and Gigabit Ethernet chipsets.  His current research interests include 5G technologies, Internet-of-Things, communications security, and locationing technologies. Dr. Wang has over 300 peer-reviewed journal and conference papers, in addition to 26 granted and pending patents and several standard contributions.

Dr. Wang is a Fellow of IEEE and an IEEE Distinguished Lecturer. He has received many awards and recognitions, including Canada Research Chair, CRC President¡¯s Excellence Award, Canadian Federal Government Public Service Award, Ontario Early Researcher Award and five IEEE Best Paper Awards. He currently serves as an Editor/Associate Editor for IEEE Transactions on Communications, IEEE Transactions on Broadcasting, and IEEE Transactions on Vehicular Technology and He was also an Associate Editor for IEEE Transactions on Wireless Communications between 2007 and 2011, and IEEE Wireless Communications Letters between 2011 and 2016. Dr. Wang was involved in a number of IEEE conferences including GLOBECOM, ICC, VTC, PIMRC, WCNC and CWIT, in different roles such as symposium chair, tutorial instructor, track chair, session chair and TPC co-chair.
\end{IEEEbiography}

\begin{IEEEbiography}{Arumugam Nallanathan}(S'97-M'00-SM'05-F'17) is Professor of wireless communications in the School of Electronic Engineering and Computer Science at Queen Mary University of London since September 2017. He was with the Department of Informatics at King¡¯s College London from December 2007 to August 2017, where he was Professor of wireless communications from April 2013 to August 2017. He was an Assistant Professor in the Department of Electrical and Computer Engineering, National University of Singapore from August 2000 to December 2007. His research interests include 5G Wireless Networks, Internet of Things (IoT) and Molecular Communications. He published more than 350 technical papers in scientific journals and international conferences. He is a co-recipient of the Best Paper Award presented at the IEEE International Conference on Communications 2016 (ICC2016) and IEEE International Conference on Ultra-Wideband 2007 (ICUWB 2007). He is an IEEE Distinguished Lecturer. He has been selected as a Web of Science (ISI) Highly Cited Researcher in 2016.He is an Editor for IEEE Transactions On Communicaitons and IEEE Transactions On Vehicular Technology. He was an Editor for IEEE Transactions On Wirelss Communcations (2006¨C2011), IEEE Wireless Communications Letters and IEEE Signal Processling Letters. He served as the Chair for the Signal Processing and Communication Electronics Technical Committee of IEEE Communications Society and Technical Program Chair and member of technical program committees in numerous IEEE conferences. He received the IEEE Communications Society SPCE outstanding service award 2012 and IEEE Communications Society RCC outstanding service award 2014.
\end{IEEEbiography}

\end{document}